\documentclass[runningheads]{llncs}

\usepackage{cite}
\usepackage{amsmath,amssymb,amsfonts}
\usepackage{algorithmic}
\usepackage{graphicx}
\usepackage{textcomp}
\usepackage{xcolor}
\usepackage{comment}
\usepackage{url}
\usepackage{booktabs}
\usepackage{float} 
\usepackage{algorithm}
\usepackage{algorithmic}
\usepackage{multirow}
\usepackage[caption=false, font=footnotesize]{subfig}
\def\BibTeX{{\rm B\kern-.05em{\sc i\kern-.025em b}\kern-.08em
    T\kern-.1667em\lower.7ex\hbox{E}\kern-.125emX}}
\title{Direct Raw Audio Signal Processing via Reservoir Computing: An Investigation into "Feature-Free" Architectures}
\author{Rinku Sebastian\inst{1}\orcidID{0000-0002-0813-8202} \and
Simon O'Keefe\inst{2}\orcidID{0000-0001-5957-2474} \and
Martin A Trefzer\inst{1}\orcidID{0000-0002-6196-6832}}
\authorrunning{R. Sebastian et al.}
\institute{School of PET, University of York, UK \and
Department of Computer Science, University of York, UK
\email {rinku.sebastian@york.ac.uk}, {simon.okeefe@york.ac.uk}, {martin.trefzer@york.ac.uk}}
\institute{}
\begin{document}

\maketitle

\begin{abstract}
This paper evaluates Reservoir Computing (RC) as an autonomous, "feature-free" framework for audio processing, designed to eliminate traditional, handcrafted feature extraction stages. We investigate whether the high-dimensional temporal dynamics inherent in a Reservoir can function as a robust end-to-end processor for the direct classification of raw acoustic signals. By bypassing computationally intensive representations like MFCCs, this approach seeks to mitigate significant intellectual and pre-processing bottlenecks in traditional signal pipelines. Our study evaluates and compares shallow, sequential, and parallel deep Reservoir architectures to determine their capacity for hierarchical feature representation. Experimental results demonstrate that the proposed parallel approach consistently outperforms shallow and sequential baselines while maintaining low model complexity. These findings highlight the potential of RC as an efficient and scalable alternative for time-domain audio processing, offering a promising pathway toward deployable, low-power acoustic systems with minimal preprocessing requirements.
\end{abstract}


\section{introduction}
Traditional audio pattern recognition relies heavily on techniques such as Mel-Frequency Cepstral Coefficients (MFCCs). These methods serve two primary tasks: reducing the high dimensionality of raw audio and highlighting perceptually relevant information. However, this dependence on handcrafted features creates a significant intellectual and computational bottleneck in traditional pipelines. This study seeks to leverage the temporal dynamics and high-dimensional projection of RC to model a ``feature-free'' framework, that is, audio processing without separate feature extraction stage. We investigate whether the Reservoir’s inherent dynamics can classify acoustic signals without any dedicated feature extraction stage, and thus simplifying the system into a single-stage, end-to-end processor.

 The concept of end-to-end speech processing~\cite{Graves2014-le} has gained prominence, where models attempt to learn the feature representation directly from the raw audio waveform, bypassing all conventional pre-processing. Early attempts showed promise and has confirmed that features learned directly from the waveform can outperform feature extracted models(like MFCCs), particularly in multi-task scenarios~\cite{Ravanelli2018}. However, this shift primarily moves the computational burden rather than eliminating it; the networks required to learn these complex transformations from raw data are often massive, demanding orders of magnitude more training data, parameters, and energy than traditional systems~\cite{Amodei2016}. Our work attempts to bridge this gap by offering the simplicity and efficiency of the RC framework while overcoming the significant temporal and computational constraints of conventional feature dependence. Unlike prior end-to-end raw waveform models based on deep CNNs, we investigate whether lightweight Reservoir computing can achieve comparable feature learning without complex processing and large number of neurons.

\section{Background}
Reservoir Computing (RC) originates from Recurrent Neural Networks (RNNs), which process temporal data via evolving internal states \cite{rumelhart1986learning}. Unlike traditional RNNs that suffer from expensive back-propagation and vanishing gradients~\cite{hochreiter1991untersuchungen}, RC utilizes a fixed, untrained high-dimensional dynamical system as a Reservoir \cite{jaeger2001echo}. This Reservoir performs a non-linear transformation of input data into a higher-dimensional space where linear separation is feasible \cite{maass2002real}. RC’s efficiency, scalability, and biological plausibility \cite{lukosevicius2009reservoir, stepney2005journey} make it ideal for real-time processing of high-dimensional temporal signals \cite{Picco2025-tb}, offering a low-latency pathway for audio classification \cite{Sebastian2026-eo}.

While traditional Echo State Networks (ESNs) use a single shallow Reservoir, Deep RC employs hierarchical stacks to capture multi-scale dynamics \cite{Ma2023}. In this structure, lower layers extract high-frequency local features while deeper layers model abstract, long-term temporal structures \cite{Gallicchio2021-yy}. Despite these advancements, most Deep RC applications still rely on handcrafted features like Mel-Frequency Cepstral Coefficients (MFCCs). This work diverges from such approaches by investigating Deep RC as an end-to-end, "feature-free" processor. By processing raw acoustic waveforms directly and training only the readout layer, we aim to bypass traditional pre-processing bottlenecks with a lightweight, computationally efficient alternative.

\section{Exploring Single RC for Raw Audio Processing}

\subsection{Data Dimensionality and Pre-Processing}
Directly processing high-resolution raw audio is computationally impractical due to the extreme length and redundancy of the time-series data, which can obscure the Reservoir’s ability to extract relevant information. To maintain a "feature-free" architecture without the overhead of spectral analysis and maintaining computational simplicity, we implement a lightweight decimation pipeline for dimensionality reduction~\cite{sebastian2025audiosignalprocessingusing}. The raw signal is first segmented into fixed-length frames of $N = 250$ samples—a window size determined empirically to optimize classification performance. Within each window, non-linear peak-to-peak detection selects a single representative value to highlight fundamental envelopes and transient features. These near-instantaneous steps reduce data volume while preserving the essential temporal and energy profiles, allowing the Reservoir to function as both an autonomous feature extractor and classifier.

\subsection{Performance Measurement and Visualization}

\begin{figure}[h!]
    \centering
    \subfloat[Digit Recognition performance]
    {\includegraphics[width=0.49\linewidth]{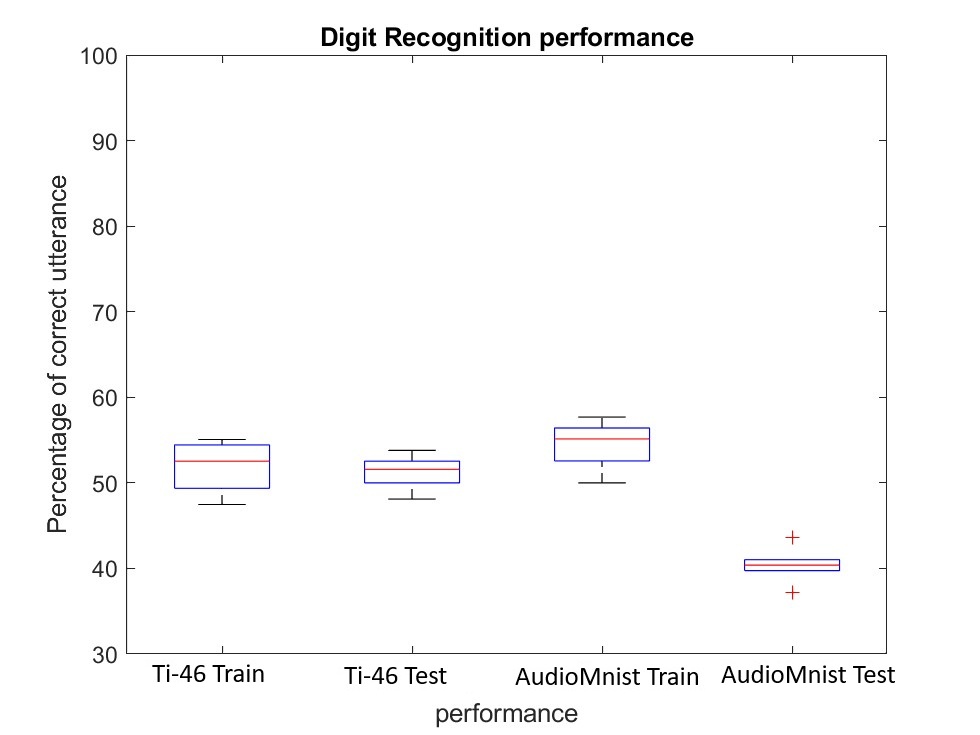}}
     \subfloat[Speaker Recognition performance]
     {\includegraphics[width=0.49\columnwidth]{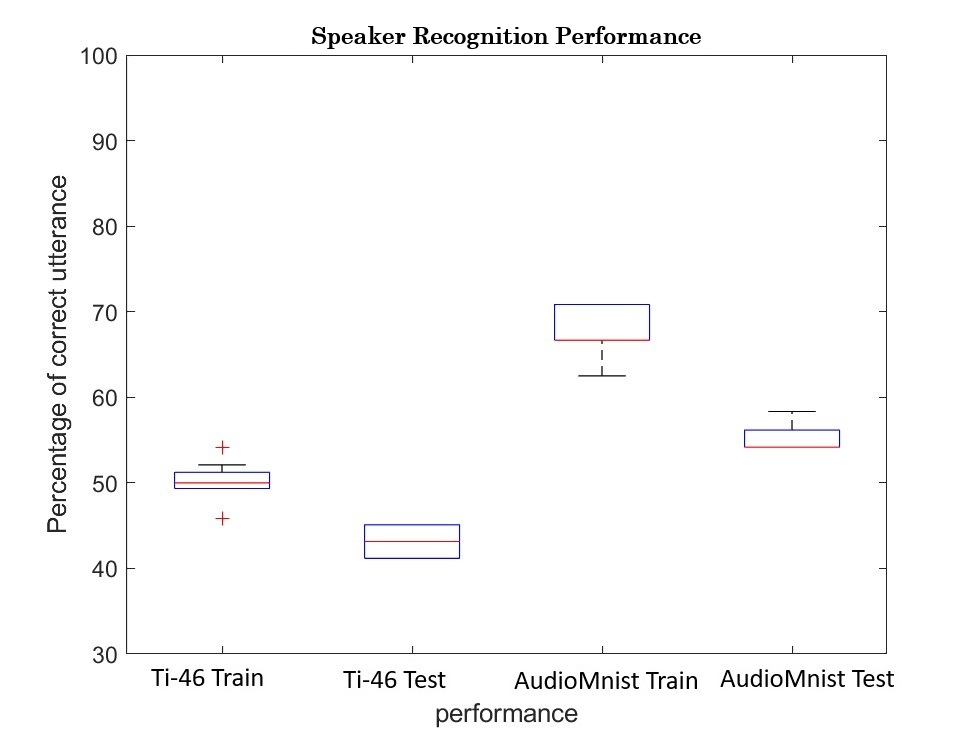}}
    \caption{Digit and speaker recognition using "feature-free" audio processing using shallow Reservoir}
    \label{fig:direct}
\end{figure}
The capacity of a single Reservoir to process audio utilising the raw audio signal without any feature extraction is demonstrated by the box plots \ref{fig:direct}.

\subsubsection{Challenges of Raw Signal Processing}
Directly feeding "feature-free" raw audio into a standard Echo State Network (ESN) is suboptimal due to two primary constraints. First, a standard ESN has a fixed timescale, making it difficult to process the multi-scale temporal events, such as transient phonemes and slower syllabic envelopes, inherent in speech. Second, a single Reservoir lacks inherent frequency decomposition; without the spectral analysis typically provided by the biological cochlea or MFCCs, the network struggles to distinguish the frequency-based patterns essential for accurate audio classification.

\subsection{From Shallow to Deep RC  Raw Audio Processing}
In seeking to maximize the Reservoir Computing (RC) system's capacity for ``feature-free'', end-to-end raw audio understanding, we extend our investigation to include the necessity and potential utility of a Deep Reservoir Computing(DRC) approach. While the canonical RC model, the Echo State Network (ESN), is inherently defined by its architectural simplicity—consisting of a single recurrent layer (the Reservoir) with fixed, random internal weights—the sheer complexity involved in transforming a lengthy, high-resolution, unprocessed audio signal into a low-dimensional, discriminative, and class-separable state space suggested that the computational capacity of a single layer might prove insufficient for the task.

\section{Sequential Deep Reservoir Architectures}
This experiment investigates a Series-Connected Deep Reservoir Computing (DRC) architecture, a multi-layered framework engineered to enhance the precision of  audio recognition without feature extraction stage. By moving beyond the limitations of a single-layer Reservoir, this design implements a hierarchical strategy for temporal feature extraction~\cite{gallicchio2017deep}. 

\begin{figure}
    \centering
    \includegraphics[width=.7\linewidth]{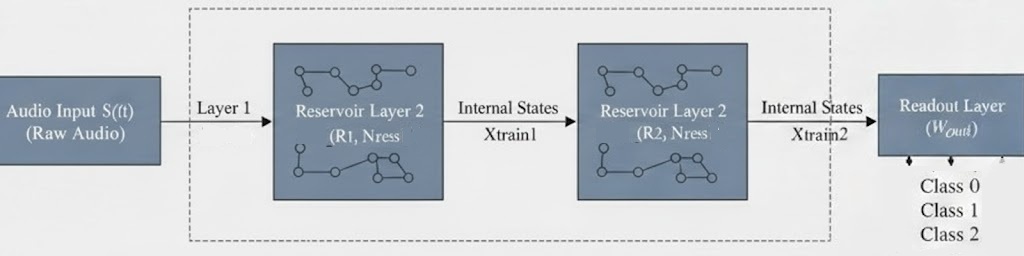}
    \caption{Deep Series "Feature-free" Audio signal processing}
    \label{fig:DSD}
\end{figure}

The architectural process begins with the primary Reservoir, which receives the raw acoustic input and maps it into a high-dimensional space. This first stage captures the immediate temporal dynamics of the signal. Rather than extracting a final classification at this point, the system harvests the internal hidden states of this primary Reservoir. These states, which serve as a filtered representation of the initial audio, are then passed as a continuous input stream into a secondary Reservoir.  Figure \ref{fig:DSD} shows the deep series architecture.

By processing these pre-encoded states, the secondary Reservoir is able to perform higher-order integration. Finally, the output of this secondary layer is fed into a trained readout mechanism for audio identification. This sequential flow effectively increases the system's fading memory and computational depth.

As illustrated in Figure \ref{fig:DSD} , the architecture follows a linear series configuration:

\begin{itemize}
    \item \textbf{Primary Input:} Raw audio signals, represented as $S(t)$, are fed into the first layer.
    \item \textbf{Reservoir Layer 1: } A Reservoir consisting of 150 neurons. It processes the audio input to generate high-dimensional internal states.
    \item \textbf{Second Layer Input:} The internal state vectors generated by first Reservoir (Xtrain1) are passed directly as the input to the subsequent layer.
    \item \textbf{Reservoir Layer 2:} A secondary Reservoir consisting of 400 neurons, that further transforms the temporal features provided by the first layer.
    \item \textbf{Readout Layer ($W_{out}$):} The internal states of the second Reservoir (Xtrain2) are used to train the readout layer, mapping the deep temporal features to the final digit/speaker classification.
\end{itemize}

\subsection{Results for Sequential Deep RCs}
The training and test performance for digit and speaker recognition on the Ti-46 dataset using series deep "feature-free" method is shown in box-plots 
\begin{figure}[h]
    \centering
   \subfloat[Digit Recognition performance]
    {\includegraphics[width=0.5\linewidth]{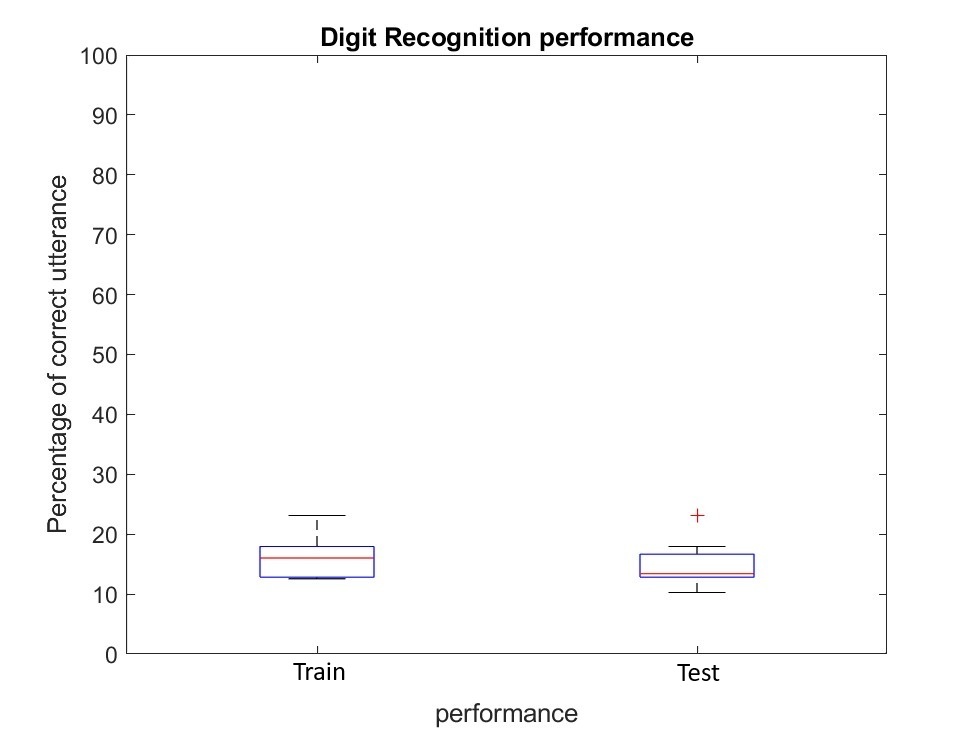}}
    \subfloat[Speaker Recognition performance]
     {\includegraphics[width=0.5\columnwidth]{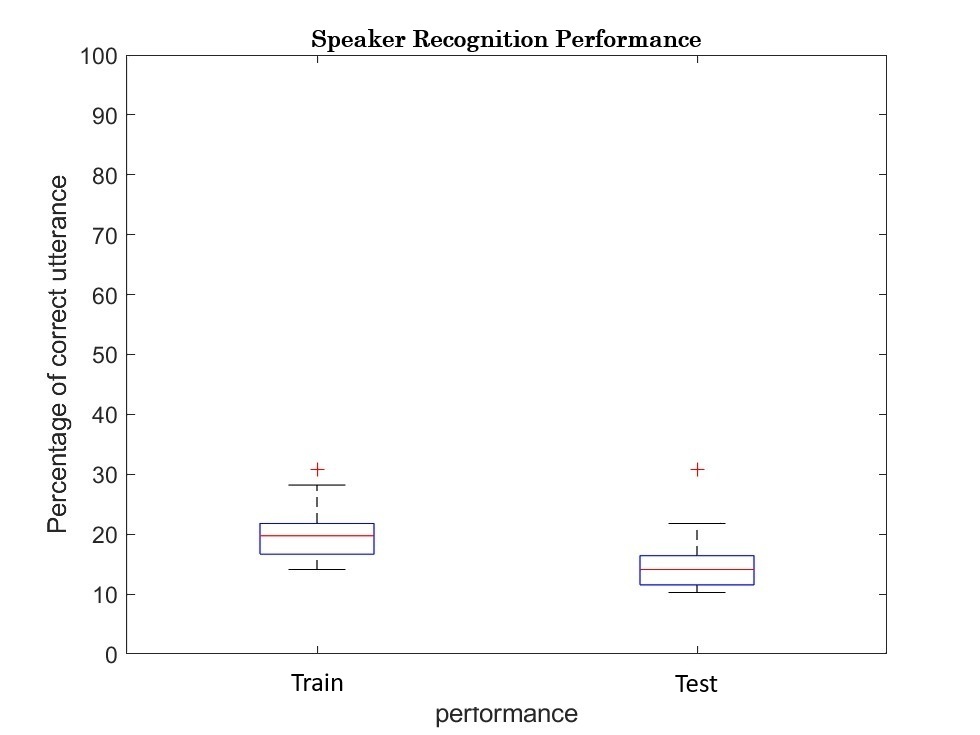}}
            \caption{Performance of series deep Reservoir used for "feature-free" audio processing}
    \label{fig:DeepSD}
\end{figure}

This experiment serves as a preliminary evaluation of the series-connected deep Reservoir architecture for "feature-free" audio signal processing. However, the performance metrics observed during testing did not meet the necessary benchmarks for viability. Consequently, the data is insufficient to support any definitive inferences or broader conclusions regarding the efficacy of this specific configuration at this time, instead as an indication that the current model requires fundamental realignment.

However, the performance discrepancies observed reveals a fundamental trade-off between feature extraction and signal degradation. When using a series-deep connection, the first Reservoir functions as a high-dimensional non-linear filter, extracting low-level acoustic features from the raw audio. However, this creates an information bottleneck; by the time the data reaches the second Reservoir, it is no longer the original signal but rather the internal state of the first layer. If hyper-parameters-like the spectral radius or leak rate-are not perfectly tuned, this first stage washes out critical high-frequency details. This signal washing forces the second Reservoir to learn from blurry data, explaining the poor accuracy.

The shallow model highlights the inherent difficulty of processing raw audio without pre-processing. Typically, audio signals are converted into Mel Frequency Cepstral Coefficients (MFCCs) to simplify the task. Without this assistance, a single Reservoir must perform the heavy lifting of frequency analysis itself. 
\begin{figure}[h!]
    \centering
    \includegraphics[width=0.65\linewidth]{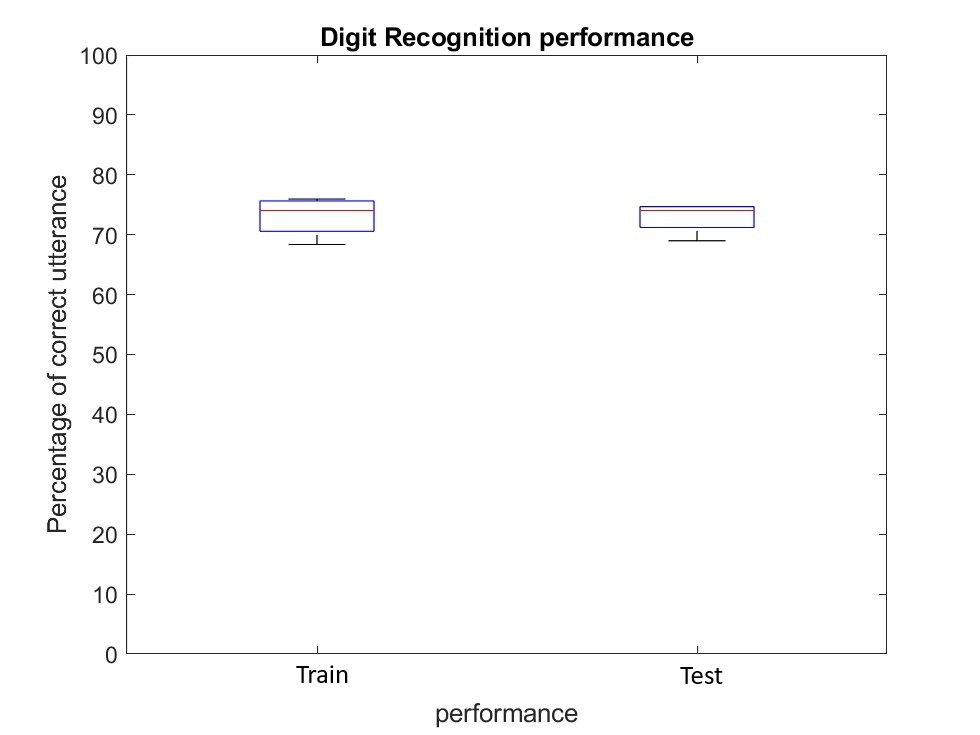}
    \caption{Performance of Deep series RC with MFCC as input}
    \label{fig:deepmfccinput}
\end{figure}
When a Deep Series Reservoir is supplied with Mel Frequency Cepstral Coefficients (MFCCs) as input rather than raw audio, it typically yields better classification results as shown in figure \ref{fig:deepmfccinput}, because the MFCCs serve as a pre-conditioned, low-dimensional representation of the acoustic manifold.

\begin{figure}[h!]
    \centering
    \includegraphics[width=.7\linewidth]{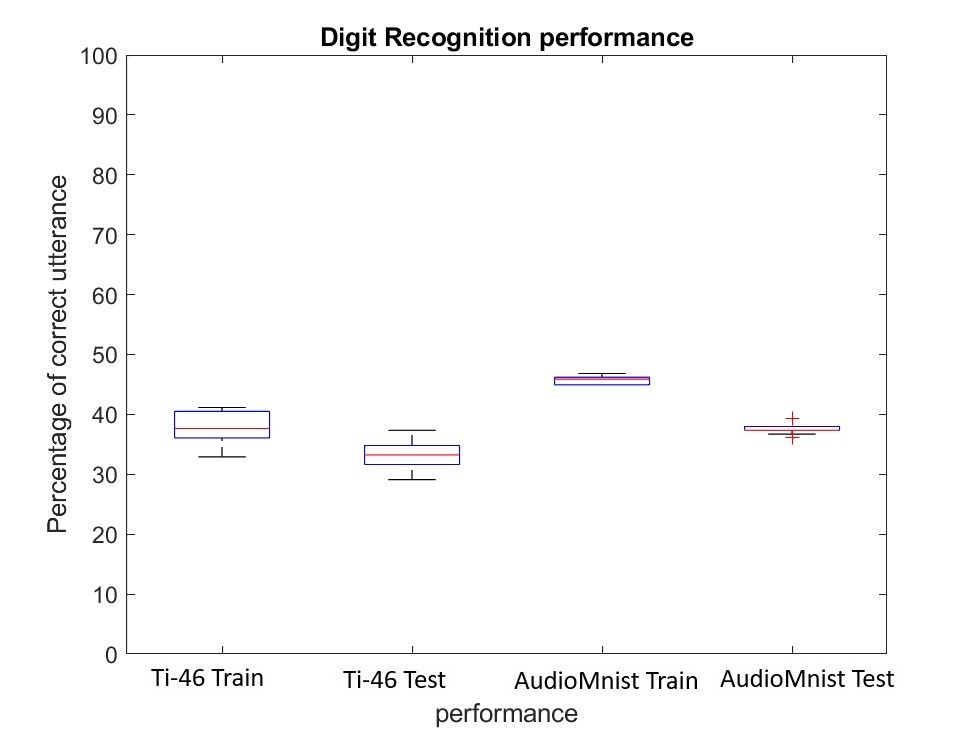}
    \caption{Output of Series Deep "feature-free" with second layer receiving a copy of Input}
    \label{fig:seriesdeepU}
\end{figure}

However, even in these experiments, the performance of the Deep Series Reservoir is lower compared to the performance of a shallow Reservoir fed with MFCCs~\cite{sebastian2026bridgingbiologicalhearingneuromorphic}. This discrepancy suggests that the second Reservoir may be amplifying errors or losing critical information if it does not receive a direct copy of the input signal from the first layer. To rectify this, one potential fix is to provide the second Reservoir with a copy of the input to maintain signal integrity throughout the hierarchy. Based on this hypothesis, we modified the Deep Series Reservoir by providing the second layer with a direct copy of the input signal alongside the output from the first Reservoir. While this architectural adjustment yielded a noticeable improvement in performance as shown in figure \ref{fig:seriesdeepU}—confirming that maintaining signal integrity is vital for deeper hierarchies—the results did not surpass the baseline shallow model. This persistent gap suggests that the raw audio signal, even when processed through multiple Reservoir stages, lacks the discriminative power of hand-engineered spectral features. 

\section{Parallel "feature-free" Reservoir Architectures} 
The Parallel "feature-free" Reservoir Computing (PFRC) approach can be interpreted as a multi-resolution ensemble, simultaneously recording many temporal perspectives of the same raw data, whereas the sequential (series) approach imitates the hierarchical abstraction~\cite{sun2024deep}. In this setup, both Reservoirs receive the raw input signal simultaneously rather than in a sequential chain. By decoupling the second layer from the transformed output of the first, we ensure that the second Reservoir is not forced to process potentially degraded or overly filtered information. This parallel approach allows each Reservoir to act as an independently —potentially operating at different temporal scales—thereby preserving the richness of the raw audio across the entire hidden state and preventing the error propagation inherent in deep serial hierarchies. It underscores why moving to a parallel strategy is more effective. It allows multiple independent Reservoirs to be tuned to capture different aspects of the signal, providing a more robust analysis than a single layer~\cite{vlachas2022parallel}.

In the parallel "feature-free" architecture, the implementation of heterogeneous leak rates facilitates a multi-scale temporal analysis of the raw audio signal, which is essential for capturing the complex dynamics of speaker-specific identities. By assigning a lower leak rate ($\alpha_{low}$) to one Reservoir, the system establishes a slow integrator with an extended fading memory. This Reservoir effectively smooths the high-frequency fluctuations of the raw waveform to capture global acoustic characteristics, such as the speaker’s rhythm and long-term resonance.~\cite{5966352}

Simultaneously, the second Reservoir, configured with a higher leak rate ($\alpha_{high}$), serves as a fast tracker. This layer is more sensitive to the immediate temporal changes in the input signal, allowing the network to preserve the fine-grained textures of speech, such as rapid consonantal attacks and pitch variations. Because these Reservoirs operate in parallel, the final readout layer can simultaneously leverage both the long-term temporal context and the immediate acoustic transients. This functional diversity prevents the information redundancy typically found in shallow models and avoids the catastrophic signal degradation observed in series-connected architectures, thereby providing a more robust and comprehensive representation of the speaker’s voice directly from the time-domain input.

By diversifying $\alpha_i$ across the parallel layers, the model ensures that the concatenated state vector $X(t) = [x_1(t); x_2(t)]$ presented to the readout layer contains a multi-resolution profile of the acoustic signal, capturing both instantaneous and evolutionary vocal features.

\begin{figure}
    \centering
    \includegraphics[width=.8\linewidth]{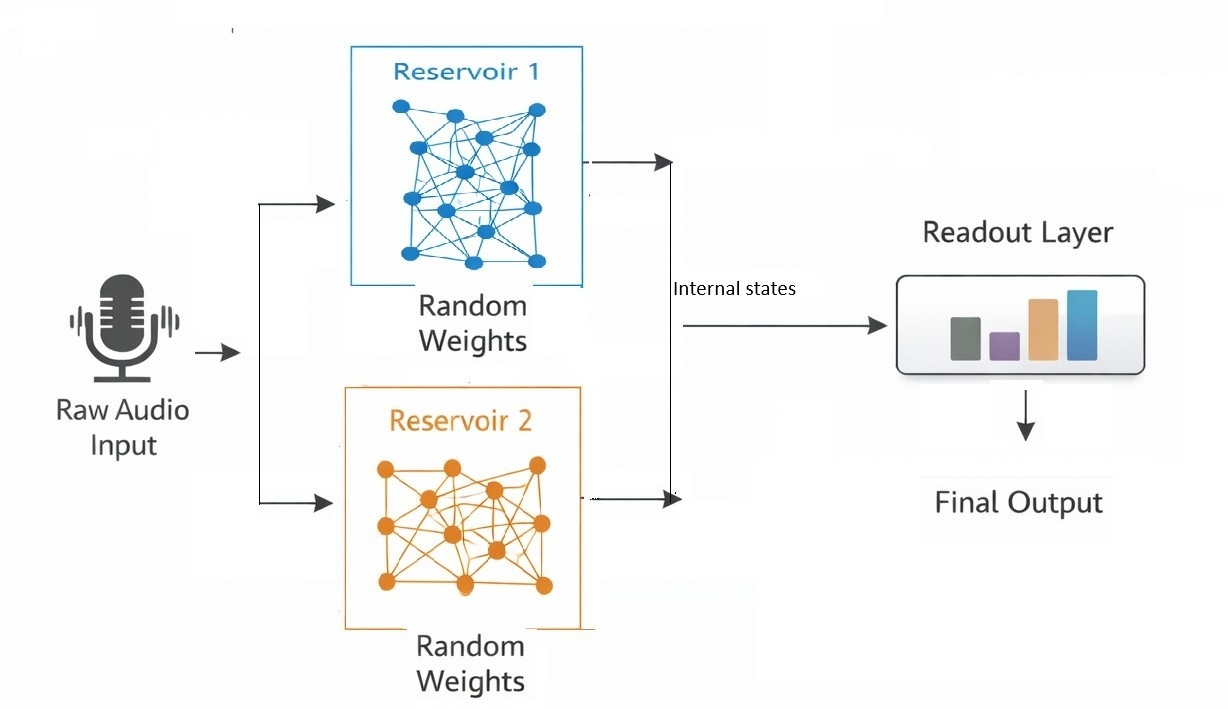}
    \caption{Parallel "feature-free" architecture}
    \label{fig:parallel}
\end{figure}

The parallel architecture achieves improved success rate by avoiding the single point of failure inherent in series models. In a parallel setup, every Reservoir receives the raw audio signal directly and simultaneously. Because each Reservoir is initialized with different random weights, they develop feature diversity, each noticing distinct nuances of the speaker's voice. This creates a redundancy where if one Reservoir misses a specific frequency cue, another likely captures it, allowing the final readout layer to aggregate the most relevant information from all paths. This approach effectively bypasses the bottleneck problem, resulting in an improvement over the shallow model. Figure \ref{fig:parallel} shows the parallel architecture.

\subsection{Architectural Advantages of Parallel Models}
Unlike sequential models that aim for depth-based abstraction, the parallel "feature-free" architecture utilizes breadth to capture a multi-scale representation of raw audio by feeding the signal into several distinct Reservoirs simultaneously. This design offers three core advantages: (1) \textbf{Multi-Scale Integration}, where varying leaking rates and spectral radii allow separate Reservoirs to track both rapid phonemic transitions and slower syllabic envelopes; (2) \textbf{State-Space Richness}, as diverse random weight initializations provide a vast library of non-linear projections that increase linear separability at the readout; and (3) \textbf{Signal Integrity}, which ensures each Reservoir maintains unfiltered access to the original input, preventing the information decay or cumulative distortion often found in deep, serial stacks. By merging these specialized, high-fidelity projections, the parallel approach achieves a robust multi-resolution analysis that captures the full hierarchical complexity of acoustic data without sacrificing signal resolution.

\subsection{Implementation Challenges in Parallel RC}
While parallel architectures avoid the propagation instabilities of sequential stacks, they introduce specific implementation hurdles. First, the \textbf{dimensionality and Computational Overhead} is a primary concern; concatenating states from multiple Reservoirs increases the output weight matrix size, causing training costs to scale cubically with the total node count. Second, \textbf{diversity enforcement} is critical to avoid state redundancy; the system requires a nuanced hyper-parameter search (spectral radius, input scaling, and sparsity) to ensure each branch produces complementary rather than correlated dynamics. Finally, \textbf{feature normalization} is necessary to prevent high-amplitude activations in one Reservoir from overshadowing subtle information in another, requiring careful scaling or regularization to ensure the linear readout effectively integrates contributions from all parallel branches

\subsection{Performance Measurement and Visualization}
\begin{figure}[h!]
    \centering
    \subfloat[Digit Recognition performance]
    {\includegraphics[width=0.50\linewidth]{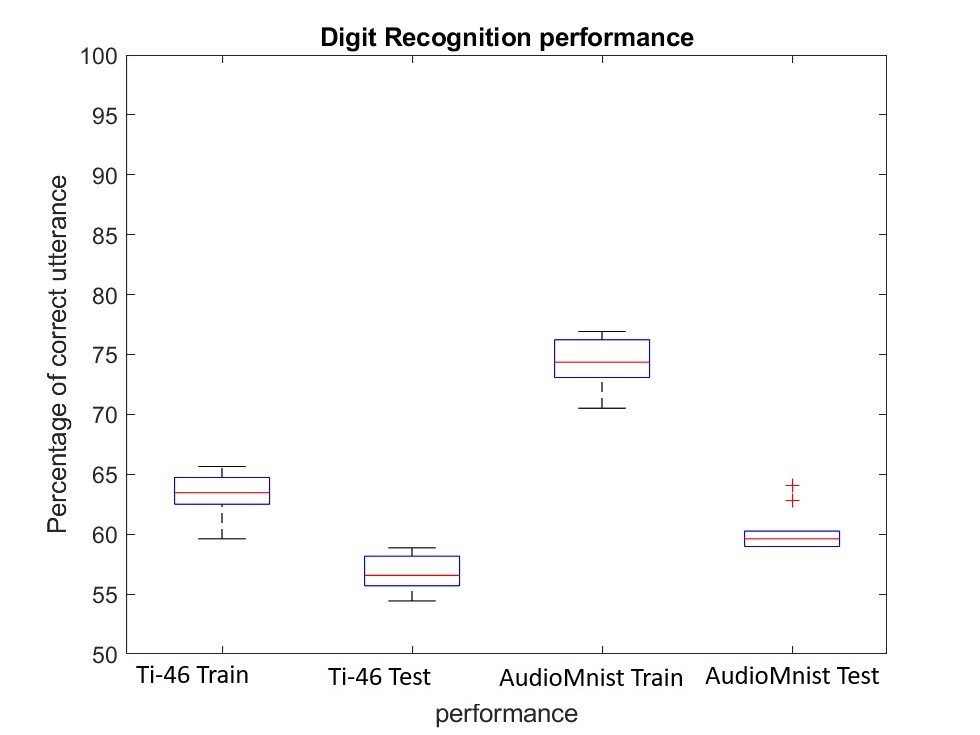}}
     \subfloat[Speaker Recognition performance]
     {\includegraphics[width=0.50\columnwidth]{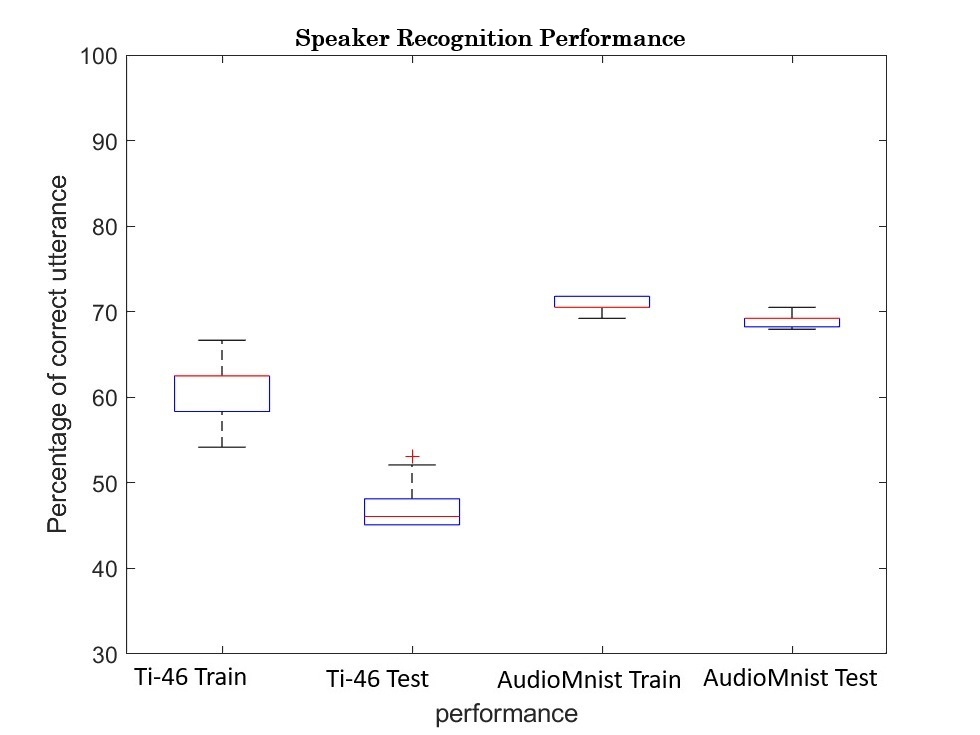}}
    \caption{Digit and speaker recognition using deep "feature-free" parallel audio processing using Reservoir}
    \label{fig:deep}
\end{figure}

The box plots \ref{fig:deep} show how a deep Reservoir may process audio using the raw audio data without any feature extraction. It is important to note that since the classification task involves 10 distinct classes, the baseline for a random guess is defined at 10\% 

\begin{table}[h!]
\centering
\resizebox{\textwidth}{!}{
\begin{tabular}{@{}lcccccccc@{}}
\toprule
\multirow{3}{2cm}{Experiments} & \multicolumn{4}{c}{Digit Recognition} & \multicolumn{4}{c}{Speaker Recognition} \\
\cmidrule(lr){2-5} \cmidrule(lr){6-9}
 & \multicolumn{2}{c}{Ti-46} & \multicolumn{2}{c}{AudioMnist} & \multicolumn{2}{c}{Ti-46} & \multicolumn{2}{c}{AudioMnist} \\
\cmidrule(lr){2-3} \cmidrule(lr){4-5} \cmidrule(lr){6-7} \cmidrule(lr){8-9}
 & Train & Test & Train & Test & Train & Test & Train & Test \\
\midrule
Shallow feature-free Audio processing & 55.06 & 53.79 & 57.69 & 43.59 & 45.09 & 41.18 & 70.83 & 58.33 \\
Parallel feature-free Audio processing & 65.65 & 58.86 & 76.92 & 64.10 & 66.67 & 53.14 & 71.79 & 70.51 \\
\bottomrule
\end{tabular}}
\caption{Comparison of different methods for Digit and Speaker Recognition}
\label{Feature-free methods}
\end{table}

Table \ref{Feature-free methods} summarises the result. 
The architectural shift from shallow to deep configurations reveals that deep Reservoirs provide a more robust framework for capturing diverse acoustic features. By distributing raw audio across independent Reservoirs simultaneously, Parallel "feature-free" Reservoir Computing (PFRC) bypasses the information bottlenecks and tedious fine-tuning inherent in stacked layers. By providing a copy of the original input signal to each subsequent layer, we prevent the loss of information in series architecture.

\section{Discussion}

The architectural evolution from shallow to deep configurations reveals a critical trade-off between structural complexity and signal integrity. A shallow Reservoir serves as the initial baseline; however, without the aid of pre-processing, a single-layer system lacks the precision required for complex frequency analysis. While Deep Reservoir Computing (DeepRC) theoretically offers a hierarchy of temporal representations—where layers capture a spectrum of fast transients and slow global patterns—the implementation method significantly dictates success.

In our experiments, a series-deep connection yielded underwhelming results due to the phenomenon of signal washing. In this configuration, the first Reservoir acts as a high-dimensional non-linear filter that often obscures high-frequency details. Consequently, subsequent layers receive a blurred internal state rather than the original audio features. 

The parallel model bypasses the information bottlenecks inherent in serial chains. By feeding the raw audio to multiple independent Reservoirs simultaneously, the architecture fosters feature diversity. This redundancy ensures that the readout layer can aggregate a broad spectrum of acoustic cues, capturing nuances that a single or stacked Reservoir might overlook. The parallel Reservoir architecture offers notable robustness and feature diversity. By eliminating the computationally expensive feature extraction stages typically required for audio processing, this "feature-free" Reservoir architecture significantly reduces the system's overall footprint. By operating entirely in the time domain with significantly fewer parameters, it stands as  lightweight model, making it ideal for low-power neuromorphic and edge computing hardware. While traditional feature based pipelines may achieve higher peak accuracy, they are difficult to implement on such hardware; in contrast, this method maintains a strategic balance between simplicity and classification effectiveness across multiple datasets. Ultimately, the parallel design captures rich, multi-scale signal details—from fast transients to slow rhythms—providing a promising alternative for autonomous signal processing.

Despite these gains, we observed a persistent challenge regarding model instability: small variations in weight initialization or hyper-parameter drift can trigger unpredictable fluctuations in the Reservoirs' internal states, potentially compromising the reliability of the output.

\section{Conclusion and future work}
In short, while a single-Reservoir approach offers high computational efficiency, it often lacks the representational depth required for peak performance. Parallel Feature-free Reservoir Computing overcomes this limitation by decoupling signals across multiple pathways, delivering a comprehensive analysis of the raw audio. Ultimately, this demonstrates that Reservoir Computing can function as a self-contained processor that projects signals into a higher-dimensional space, effectively enabling audio processing without the need for traditional handcrafted features.

This framework demonstrates that Parallel Feature-free Reservoir Computing is effective for end-to-end audio processing. Although the model currently shows inconsistent performance, this sensitivity reflects the complex, high-dimensional dynamics of raw waveforms. By avoiding traditional filters, the system captures rich signal details that are typically suppressed.

Future work will focus on formalizing stabilization protocols to harness these dynamics with greater reproducibility across diverse acoustic environments. Furthermore, while the current peak-to-peak sampling method preserves structural simplicity and avoids the computational overhead of complex dimensionality reduction methods, we propose that the integration of adaptive dimensionality reduction methods will further optimize the data footprint. By aligning Reservoir temporal scales with the information density of the raw signal, the PFRC can transition from a promising experimental architecture to a robust, ultra-low-power standard for autonomous signal processing pipelines.

\bibliographystyle{plain}
\bibliography{Reference}
\end{document}